\documentclass[manuscript,screen,10pt]{acmart}
\AtBeginDocument{%
  \providecommand\BibTeX{{%
    \normalfont B\kern-0.5em{\scshape i\kern-0.25em b}\kern-0.8em\TeX}}}
\usepackage{subcaption}
\usepackage[linesnumbered,ruled]{algorithm2e}
\usepackage{tabularx,lipsum,environ,amsmath}
\usepackage{graphicx}
\usepackage{etoolbox}
\usepackage{xcolor}
\usepackage{physics}
\newcommand{\etal}{\emph{et al.\ }}
\usepackage{multirow}
\usepackage{braket}
\usepackage{caption}
\usepackage{color,xspace,listings,comment}
\usepackage[colorinlistoftodos]{todonotes}
 \renewcommand{\footnote}{\sidenote}
\begin{document}
\title{A Synergistic Compilation Workflow for Tackling Crosstalk in Quantum Machines}

\author{Fei Hua}
\email{huafei90@gmail.com}
\affiliation{%
  \institution{Rutgers University}
  \country{USA}
}

\author{Yuwei Jin}
\email{yj243@scarletmail.rutgers.edu}
\affiliation{%
  \institution{Rutgers University}
  \country{USA}
}

\author{Ang Li}
\email{ang.li@pnnl.gov}
\affiliation{%
  \institution{Pacific Northwest National Laboratory}
 \country{USA}
}

\author{Yanhao Chen}
\email{chenyh64@gmail.com}
\affiliation{%
  \institution{Rutgers University}
  \country{USA}
}

\author{Chi Zhang}
\email{raymond.chizhang@gmail.com}
\affiliation{%
  \institution{University of Pittsburgh}
  \country{USA}
}

\author{Ari Hayes}
\email{arihayes@gmail.com}
\affiliation{%
  \institution{Rutgers University}
  \country{USA}
}

\author{Hang Gao}
\email{hg405@scarletmail.rutgers.edu}
\affiliation{%
  \institution{Rutgers University}
  \country{USA}
}

\author{Eddy Z. Zhang}
\email{eddy.zhengzhang@gmail.com}
\affiliation{%
  \institution{Rutgers University}
  \country{USA}
}

% \affiliation{%
%   \institution{Rutgers university}
% %   \streetaddress{P.O. Box 1212}
% %   \city{Dublin}
% %   \state{Ohio}
% %   \country{USA}
% %   \postcode{43017-6221}
% }

\begin{abstract}
  Near-term quantum systems tend to be noisy. Crosstalk noise has been recognized as one of several major types of noises in superconducting Noisy Intermediate-Scale Quantum (NISQ) devices. Crosstalk arises from the concurrent execution of two-qubit gates, such as \texttt{CX}, on nearby qubits. It might significantly raise the error rate of gates in comparison to running them individually. Crosstalk can be mitigated through scheduling or hardware machine tuning. Prior scientific studies, however, manage crosstalk at a really late phase in the compilation process, usually after hardware mapping is done. It may miss great opportunities of optimizing algorithm logic, routing, and crosstalk at the same time. In this paper, we push the envelop by considering all these factors simultaneously at the very early compilation stage. We propose a crosstalk-aware quantum program compilation framework called CQC that can  enhance crosstalk-mitigation while achieving satisfactory circuit depth. Moreover, we identify opportunities for translation from intermediate representation to circuit for   application-specific crosstalk mitigation, for instance, the \texttt{CX} ladder construction in variational quantum eigensolvers (VQE). Evaluations through simulation and on real IBM-Q devices show that our framework can significantly reduce the error rate by up to 6$\times$, with only $\sim$60\% circuit depth compared to state-of-the-art gate scheduling approaches. In particular for VQE, we demonstrate 49\% circuit depth reduction with 9.6\% fidelity improvement over prior art on the H4 molecule using IBMQ Guadalupe. Our CQC framework will be released on GitHub.
\end{abstract}
\maketitle

\section{Introduction}
% Current Noisy Intermediate-Scale Noisy(NISQ) computer can arrive at a solution for certain classically-intractable problems at a significantly faster speed. QC has proven effecitive in  Cybersecurity, Chemistry, Complex manufacturing and Artificial intelligence. IBM\cite{IBMQ}, Google and Intel has recently announced NISQ with 49-72 qubits and will announce even more qubits in the near term which makes application more practical.

% Although large-scale of qubits is critical to different applications, qubit states are volatile and it will be affected by different quantum noises(Readout errors, gate errors and crosstalk errors). Crosstalk error is a combination of unwanted interactions between coupled qubits on quantum circuit. Among all the errors, crosstalk error is the major errors[cite] which domain the overall fidelity of the circuit. Vast of studies has been worked on Quantum Error Correction Code(QECC) and quantum mitigation try to give a promising result.

Quantum computing are known to solve certain classically intractable problems across various domains such as cybersecurity \cite{mosca+:sp18}, chemistry \cite{kauzmann+:qcbook13, cao+:cr19}, complex manufacturing \cite{hanley+:crypto14}, and artificial intelligence \cite{dunjko+:rpp18, wichert2020principles}. IBM, Google, and Rigetti among others have delivered superconducting quantum devices with dozens of qubits through the cloud. Particularly, it is expected that IBM will deliver thousand-qubit devices soon \cite{ibmq:roadmap}, bringing quantum computing closer to technical reality.

\begin{figure}[htb]
  \centering
  \includegraphics[width=0.75\linewidth]{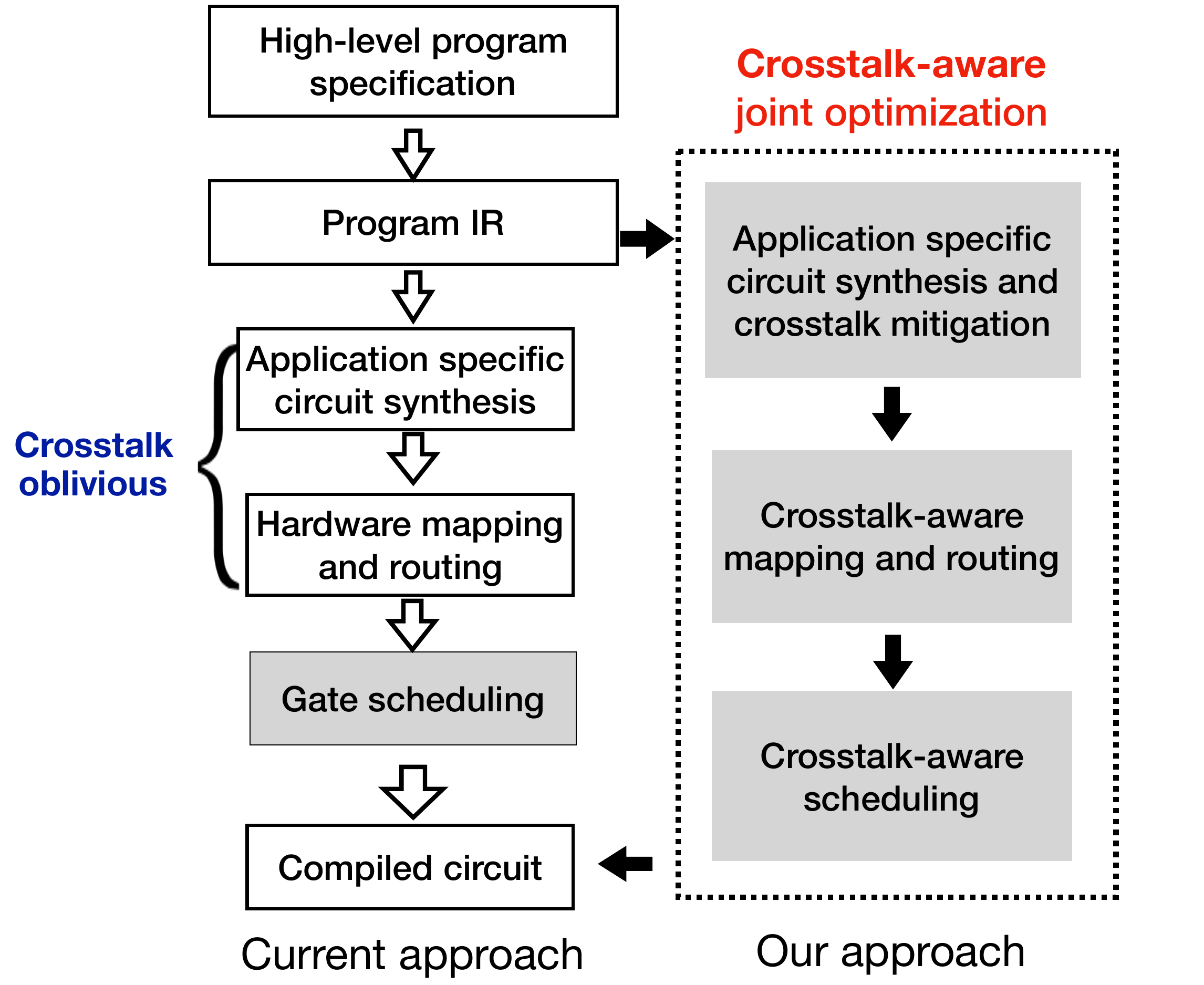}
  \caption{Our quantum program compilation workflow.}
  \label{fig:compileflow}
\end{figure}

Today's superconducting NISQ systems are prone to various types of errors: decoherence error \cite{shor+:focs94}, readout error \cite{linke+:sa17}, gate error \cite{greenbaum+:arxiv15}, and crosstalk error. In particular, the crosstalk error is reported to have significant impact, which could be 10x worse  than other errors \cite{sheldon+:pra16, heinz+:prb21}. Crosstalk error arises from unwanted coupling between concurrent qubit operations under similar frequency \cite{sheldon+:pra16}. Two gates that are on physically nearby qubits are prone to crosstalk. Running two gates that have crosstalk will lead to higher error rate for each gate than when running each of the two gates independently.  In summary, the three contributing factors to superconducting crosstalk are the following: (a) \emph{spatial closeness}; (b) \emph{temporal closeness}, and (c) \emph{similar operating frequency}.

\begin{figure*}[htp]
  \centering
  \includegraphics[width=0.8\linewidth]{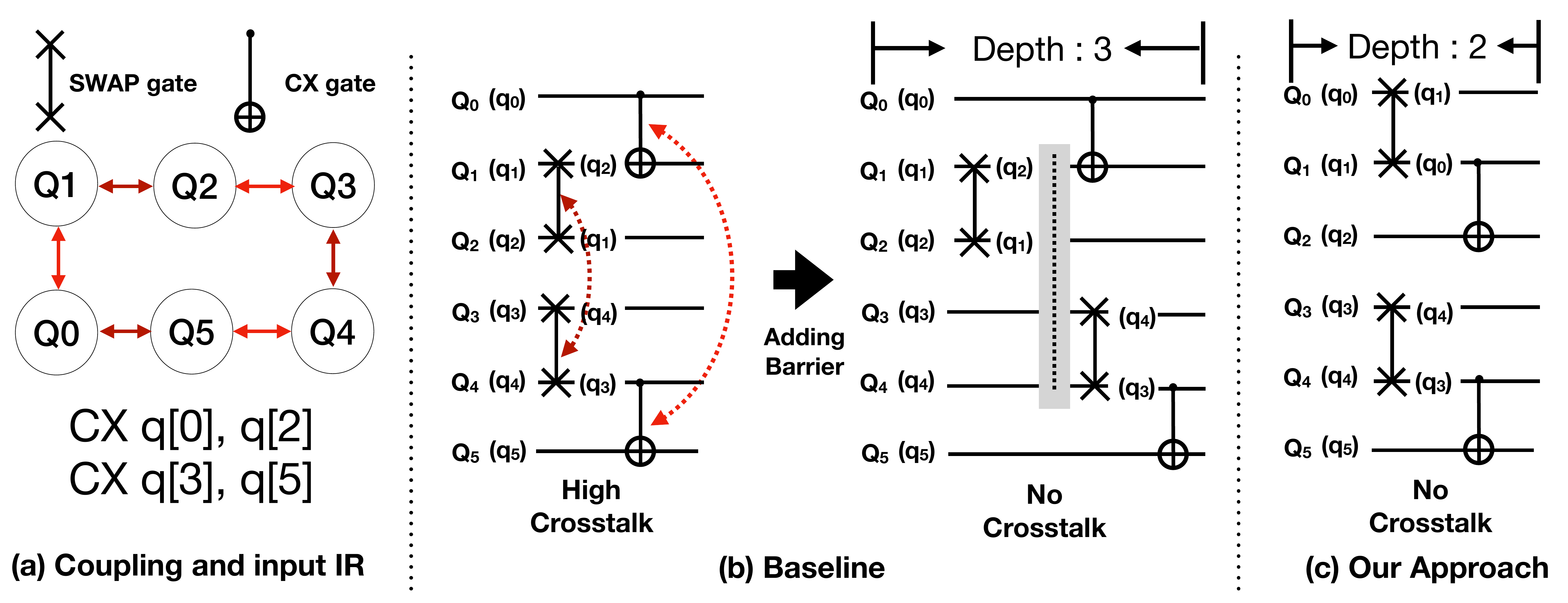}
  \caption{Crosstalk mitigated at the hardware mapping level: (a) The hardware coupling graph. Crosstalk exists between every pair of adjacent links. For instance, \emph{($Q_0$, $Q_1$)} and \emph{($Q_4$, $Q_5$)}; \emph{($Q_1$, $Q_2$)} and \emph{($Q_3$, $Q_4$)}, which implies a significantly higher error rate if two double-qubit gates are executed simultaneously than individually on these pairs. (b) The baseline approach that chooses a SWAP insertion strategy with minimal depth, but introduced high crosstalk \cite{li+:asplos19}. Then a gate must be delayed to avoid crosstalk \cite{murali+:asplos19}, causing depth to be 3. (d) Our SWAP-insertion approach that  that attains the maximum parallelism too but without any crosstalk.  Our depth is 2. Note that if we assume a SWAP takes 3 CX. Our depth is 4, while the baseline's depth is 7.  }
  \label{fig:runthroughexample}
\end{figure*} 

Recent studies have aimed to mitigate crosstalk from hardware perspective and software perspective. From hardware perspective, certain hardware devices \cite{barends+:prl19, arute+:nature19} allow dynamic frequency tuning. Hence it it can be leveraged to break one of the conditions for crosstalk \cite{ding+:micro20, klimov+:arxiv20snake}. However, it requires the hardware to have such capability. Different vendors use different technology. Not all support tunable frequency. For instance, most available quantum machines by IBM are implemented using fixed-frequency transmons.  

From software perspective, it mitigates crosstalk at compile-time, runtime, and pulse level. It  does not require hardware modification. Our work belongs to this category. Murali \etal \cite{murali+:asplos19} alleviate crosstalk at gate scheduling level, by delaying one of the two interfering gates to a later time. Delaying gates may lead to larger depth and potentially larger decoherence error. They rely on an SMT solver to  find a tradeoff between crosstalk and depth. Xie \etal \cite{xie+:asplos22} suppresses ZZ crosstalk using a co-design approach. It takes crosstalk into consideration during pulse generation stage. It leverages pulse scheduling and topology partitioning together to suppress ZZ-crosstalk for the whole circuit. 

Compared with prior software approaches \cite{murali+:asplos19, xie+:asplos22}, our approach is unique in that we take an up-the-stack approach. Rather than waiting until scheduling and pulse phase, we discover that crosstalk can already be mitigated as early as the hardware mapping and IR-to-circuit phase. To better illustrate this idea, an end-to-end quantum compiler includes the following phases: (1) Converting a high-level program specification into intermediate representation (IR), (2) converting IR into a logical circuit \cite{li+:asplos22}, (3) hardware mapping \cite{li+:asplos19, Zulehner+:DATE18, siraichi+:cgo18} from logical circuit to physical circuit, (4) gate scheduling of the physical circuit, and (5) converting gates to pulses and performing pulse scheduling/optimization \cite{xie+:asplos22}. The entire workflow converts a high level program specification into control pulses that can be recognized by the low-level hardware.

Existing software approaches \cite{murali+:asplos19, xie+:asplos22} have optimized crosstalk at the phases of gate scheduling and pulse generation/scheduling. Our major finding of this paper is that opportunities exist up the stack of the compiler flow for mitigating crosstalk. Postponing until the gate scheduling and pulse generation phases may miss these opportunities.

These opportunities for mitigating crosstalk emerge in two scenarios. First, it is the \emph{hardware mapping} stage, where SWAP gates are inserted to overcome the topology connectivity constraint. At this stage, a crosstalk-oblivious strategy might lead to a less desired circuit regardless how efficient the scheduling stage is. We show an example that at hardware mapping level, we can properly choose a right SWAP insertion strategy such that crosstalk can be completely mitigated without having to delay gates at scheduling stage, in Fig. \ref{fig:runthroughexample}.

Second, the opportunities arise at the \emph{application level}. An important class of applications in quantum computing  named variational quantum algorithms (VQA) \cite{parrish+:prl19, li+:asplos22} allow circuits to be synthesized in a flexible way from IR: (a) the order of certain gate(-block)s can be flexibly determined, and (b) 
even which gates to be synthesized is flexibly determined as long as a tree-construction constraint is met. 

\begin{figure}
    \centering
    \includegraphics[width=0.75\textwidth]{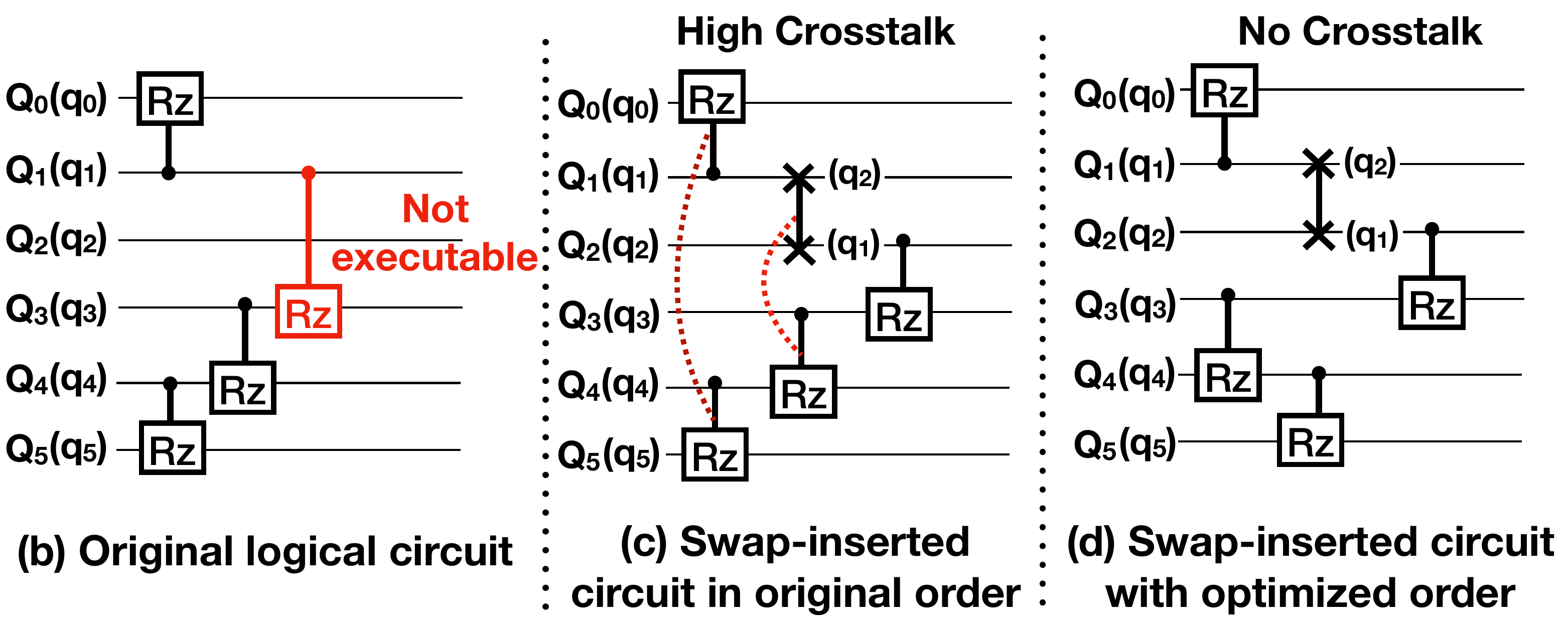}
    \caption{QAOA: Exploiting the flexibility in gate ordering. QAOA is a type of VQA algorithm. The controlled-Rz gate can commute without affecting the outcome. We use the same hardware coupling as in Fig. \ref{fig:runthroughexample} (a). Now (b) shows the original ordering of gates. (c) Executing according to the original order causes crosstalk. (d) A different order (switching Rz(q3,q4) and Rz(q4,q5) leads to no crosstalk, same depth. }
    \label{fig:qaoa_example}
\end{figure}

\begin{figure*}
    \centering
    \includegraphics[width=0.75\textwidth]{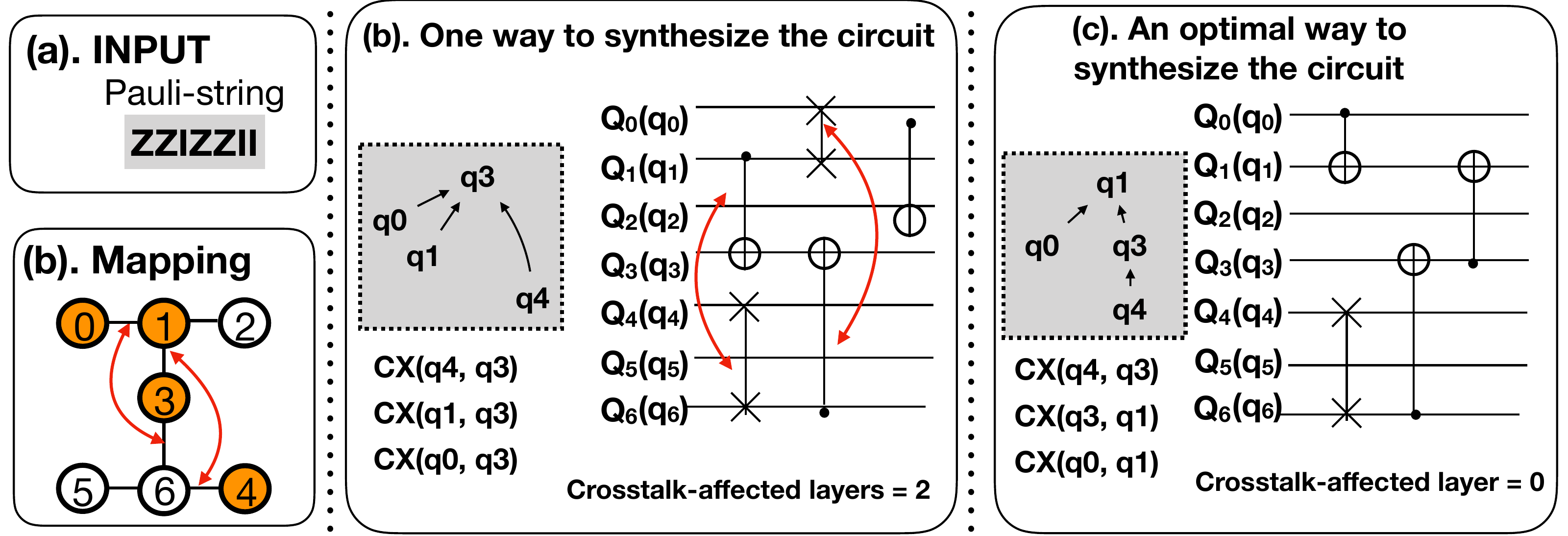}
    \caption{VQE Example: Exploiting the flexibility in gate synthesis. In this example, we focus on Hamiltonian simulation. (a) Shows the Pauli-string to be simulated. For all non-identify operators (on qubit 0, 1, 3, and 4), a circuit synthesizer must construct a directed tree where a node can follow links (implemented as CNOT gates) to the root. The execution order is also specified by the directions in the tree edges. (b) shows a synthesized tree (and corresponding circuit) that high crosstalk exists if a depth of 3 must be achieved. (c) shows an optimal way to synthesize the circuit with optimal depth is achieved without any crosstalk. Note that the other half of a synthesized circuit is omitted here since they are symmetric. }
    \label{fig:vqe_example}
\end{figure*}

As a matter of fact, VQA happen to be one of the very few applications that can demonstrate quantum advantage using near-term quantum devices \cite{alam+:dac20, alam+:micro20}. We show two motivating examples for application-specific optimization. In the first example we show a circuit where by exploiting the flexibility of the order of gates, we suppress the crosstalk without affecting the depth. It is in Fig. \ref{fig:qaoa_example}. In the next example, we show how different circuit synthesis with respect to a tree-based constraint  can help yield less crosstalk, in Fig. \ref{fig:vqe_example}.

To exploit these unique opportunities, we introduce a crosstalk-aware quantum program compilation framework, as shown in Fig. \ref{fig:compileflow}. The difference between our crosstalk-mitigation compiler and prior work is highlighted in Fig. \ref{fig:compileflow}. Compared with the existing framework that perform crosstalk-oblivious \cite{li+:asplos22} IR-to-circuit optimization and hardware mapping, we perform a cross-stack joint crosstalk-aware optimization. The application-specific component is dedicated to VQA applications. For general applications, crosstalk mitigation mainly happens at the hardware mapping and gate scheduling phases. For VQA applications, crosstalk mitigation happens as early as in the circuit construction phase.

As far as we know, our framework is the first to take an up-the-stack approach to tackle the crosstalk problem.  As other types of errors also exist in quantum hardware, for instance, decoherence errors, we also exploit a tradeoff between depth and crosstalk as that by Murual \etal \cite{murali+:asplos19} to achieve maximum fidelity. But we use a lightweight heuristic approach that does not require SMT solver. Our main contributions are as follows:

\begin{enumerate}
    \item We are the first to reveal that crosstalk can be mitigated at as early as the hardware mapping and IR-to-circuit process. Based on this finding, we build a crosstalk-aware quantum compilation framework. 
    \item Our framework can handle both generic applications and VQA applications. We added a separate layer between IR and hardware mapping for crosstalk mitigation such that any application-specific circuit synthesis  opportunities can be exploited during the compilation workflow. 
    \item We exploit the tradeoff between crosstalk and other types of errors (mainly the decoherence error), as that by Murali \etal \cite{murali+:asplos19}. We develop a simple yet effective heuristic without having to run an optimal SMT solver.   
    \item We propose a chromatic method that can model crosstalk and dependence constraint simultaneously. We seamlessly integrate it with the other components of the framework and ensure both application-specific and generic optimizations. It works efficiently while ensuring low compilation overhead.
    \item We evaluate our framework in terms of fidelity, depth, and compilation overhead for the generated circuits. The results demonstrate significant fidelity improvement (on average 20\% to 30\%) with merely half of the original circuit depth. Particularly for VQE on H-2, our approach enhances the fidelity from 26\% to 42\%.

    %not only shows the promising improvement in fidelity for an average of 20\%-30\% but also only takes near half of the circuit depths. For specific case study like VQE, we can improve the fidelity from 20\% to 40\% with only half of the depths which is a dramatically improvement.
    
    %By testing the scalability, we use different size of benchmarks and also under different quantum architecture for example Melborune and Tokyo, we also can change different crosstalk properties based on quantum machine's backend-property.
    % scalability for different size of benchmarks and also under different quantum architecture for example Melborune and Tokyo, we also can change different crosstalk properties based on quantum machine's backend-property.[TODO, adding data number on scalability, move to later]
\end{enumerate}
The remainder of the paper is organized as follows. We describe the background in \textsection\ref{sec:background}.  We present our crosstalk-aware quantum program compilation framework in \textsection\ref{sec:techoverview} and describe details in  \textsection\ref{sec:techgeneric} \textsection\ref{sec:techoverview}. We present experiment analysis in \textsection\ref{sec:eval} including depths, fidelity, and compilation time. We summarize related work in \textsection\ref{sec:related} and conclude in \textsection \ref{sec:conclusion}. 

% \fixit{Fei, I thought we had better fidelity results, i.e., improvements of fidelity of 5X or 6X, can you double check experiment results? Thanks!}

% we first introduce, build..
% performance advantage
% scalability
% commutavity special benchmark
%

\section{Background}
\label{sec:background}

\subsection{Circuit Error Model and Crosstalk}

1) \emph{Single-qubit gates and decoherence error}: In superconducting transom systems, single-qubit gates are implemented by driving the target qubit through microwave in the pulse level \cite{alexander+:qiskitpulse20}. For example, \texttt{RX} and \texttt{RY} rotations can be implemented by sending microwave voltage signals. However, due to imperfect implementation, all these gates can incur singe-qubit error. Besides, qubit state can decay in two ways: i) T1 relaxation
(i.e., spontaneous loss of energy leading to decay from excited state $\ket{1}$ to
ground state $\ket{0}$, and ii) T2 dephasing (i.e., loss of relative quantum phase
between $\ket{0}$ and $\ket{1}$). We can model the two decays using the following equation:
\begin{equation*}
q(t) = (1-e^{-t/T1} )(1-e^{-t/T2} )
\end{equation*}
where $t$ is qubit life-time; T1 and T2 are constants characterizing the speed of
the decays, for a qubit $q$.
% 2) Readout/measurement error: 

% \begin{figure}
% \begin{subfigure}
%   \centering
%   \includegraphics[width=.8\linewidth]{fig/Google.pdf}
%   \caption{ Google Sycamore (54 qubits)}
%   \label{fig:sfig1}
% \end{subfigure}
% \begin{subfigure}
%   \centering
%   \includegraphics[width=.8\linewidth]{fig/IBM.pdf}
%   \caption{IBM Montreal (27 qubits)}
%   \label{fig:sfig2}
% \end{subfigure}
% \begin{subfigure}
%   \centering
%   \includegraphics[width=.8\linewidth]{fig/Regetti.pdf}
%   \caption{Rigetti(16 qubits)}
%   \label{fig:sfig3}
% \end{subfigure}
% \caption{Near-term device topologies.}
% \label{fig:fig}
% \end{figure}

2) \emph{Two-qubit gates error and cross-talk error}: two-qubit gates
play a critical role in quantum computing, as they enforces entanglement between two qubits. Commonly used two-qubit gates include \texttt{CX} (also known as \texttt{CNOT}), \texttt{CZ} and SWAP. Two-qubit gate also has errors and tend to have higher error rate than single-qubit gates.

On top of the individual two-qubit gate error, crosstalk exists when two (pairs of) qubits are accidentally turned on (or close to) resonance,  using very close interaction frequencies. Crosstalk exists for all superconducting NISQ machines. Previous  study by Murali \etal \cite{murali+:asplos19} shows that crosstalk can degrade the error rate of a  two-qubit gate  by as much as 10x. Moreover, there is a variability of crosstalk error in the architecture. That is, some pairs of gates are more prone to crosstalk errors while others are less affected by the interference  \cite{murali+:asplos19}.

\subsection{Quantum Program Compilation Process}
From a high-level quantum program to control pulses ready for execution on quantum hardware, there are several stages. Firstly, the high-level program is translated into a quantum intermediate representation (IR). Next the IR should be further converted to a logical circuit. Depending on the application, the logical circuit may be flexibly \cite{li+:asplos22} synthesized or there is only one way to synthesize a logical circuit, in most cases.

A logical circuit needs to be converted into a physical circuit.  When applying a two-qubit gate, two participating qubits need to be physically connected. SWAP operations are inserted to move logical qubits to prepare for double-qubit execution. The hardware mapping stage addresses the SWAP insertion problem.  
Crosstalk mitigation is typically omitted in the hardware mapping stage in previous studies, as shown in Fig. \ref{fig:compileflow}.

After the hardware mapping process, it comes to the gate scheduling process.  The crosstalk-adaptive schedule by Murali \etal \cite{murali+:asplos19} chooses to delay the gates that have crosstalk if necessary. Delaying is implemented by barriers inserted to the circuit, as shown in our example in Fig. \ref{fig:runthroughexample} (b). 

Our work considers joint optimization of crosstalk through a cross-stack approach, covering IR-to-circuit, hardware mapping, and scheduling stages.

\subsection{Application-specific Optimization}

We focus on VQA for application-specific optimization. Unlike other generic quantum circuit, VQA application gives us the flexibility to synthesize or transform the quantum circuit. In the IR level specification of VQA applications \cite{li+:asplos22}, each program be expressed as a sequence of Pauli strings, where gates must be performed on qubits that have non-identify operator. An example is in Fig. \ref{fig:vqe_example} (a).

We discuss two categories of VQA applications. One special type of VQA application is the 2-local simulation/optimization applications, such as the Ising model  and the QAOA-Maxcut problem.  Each Pauli-string in a 2-local VQE application only contains two non-identity operators. This type of quantum application is quite common. In this particular class of applications, the gates do not impose any dependence among themselves \cite{lao+:arxiv21twoqan, alam+:dac20, alam+:micro20}. They can run (commute) in any order, without affecting the outcome of the circuit (in the perfect hardware case). An example is in Fig. \ref{fig:qaoa_example}.

 The other type of VQE application is n-local, meaning local interaction between n qubits. A Pauli-string may contain more than two  non-identity operators. The circuit can be constructed in any form as long as the n involved qubits are connected using CX gates while the formed graph must be a directed tree. There has to be a root. And that every qubit is able to interact with the root through the directed edges in the tree. The execution dependence order is also enforced by the tree. The gates must be executed from the leaves to the root (as if climbing a ladder in the tree) \cite{li+:asplos22}. An example of two different trees for one Pauli-string is shown in Fig. \ref{fig:vqe_example}.

Li \etal \cite{li+:asplos22} considers such flexibility when performing hardware mapping to minimize depth and gate count. Our framework is the first that exploits such flexibility in circuit synthesis for improving the fidelity and mitigating crosstalk.

\begin{figure}[htp]
  \centering
  \includegraphics[width=0.6\linewidth]{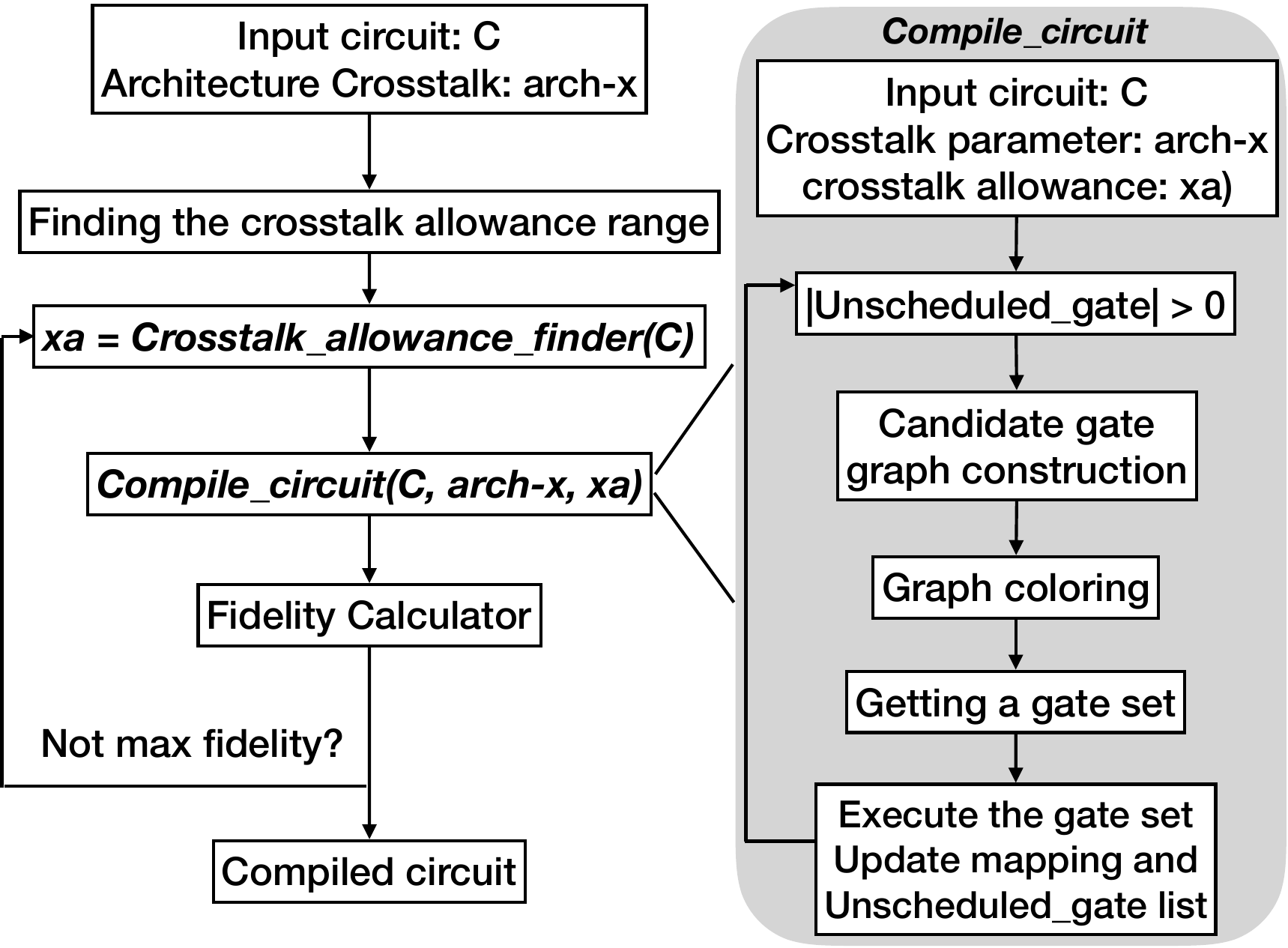}
  \caption{CQC's detailed workflow}
  \label{fig:frame}
\end{figure}

\begin{algorithm}[t!]
       \caption{Compile circuit given a crosstalk allowance xa}
       \label{alg:hs}
         \SetKwFunction{FMain}{$compile\_circuit$}
   \SetKwProg{Fn}{Function}{:}{}
  \Fn{\FMain{C, h, arch-X, xa}}{ //C is input circuit or IR, h is hardware coupling graph, arch-X is crosstalk profile, and xa is crosstalk allowance
  
        \While{$|Unsched\_gate|$ > 0}{
            V, E = \{\}, \{\}; \\
            CSG = \{V, E\};\\
            Cgates = getExecutable(Unsched\_gate, xa, h); \\
            SWAPs = getUsefulSWAP(Unsched\_gate, h); \\
            V.add(Cgates, SWAPs); \\
            depEdges = getConflict(Cgates, SWAPs); \\
            crosstalkEdges = getXtalk(h, arch-X, xa); \\
            E.add(depEdges, crosstalkEdges); \\
            top\_k\_sets = graph\_coloring(CSG);\\
            gs = top\_k\_sets.select();\\
            Unsched\_gate.update(C, gs);\\
            Qubitmapping.update();\\
            
        }
}
\end{algorithm}

{\section{Overview of CQC}}
\label{sec:techoverview}

\begin{figure*}[htp]
  \centering
  \includegraphics[width=0.9\linewidth]{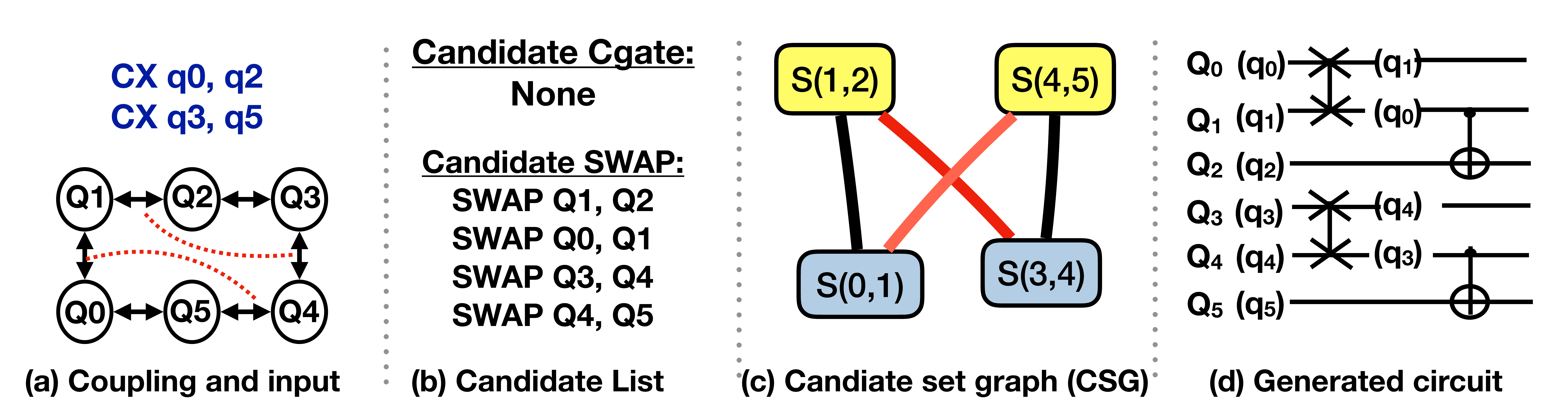}
  \caption{The Chromatic Scheme for Hardware Mapping in CQC: (a). The coupling graph and the logical gates to be executed, assuming $q_i$ maps to $Q_i$. Crosstalk is between \{Q1, Q2\} and \{Q3, Q4\}, and between \{Q0, Q1\} and \{Q4, Q5\}. (b). Candidate gate set: Since no CNOT is directly executable, the Cgate set is empty. Candidate SWAP set: There are four possible SWAP gates to move q0 and q2 closer, and q3 and q5 closer. (c). CSG constructed with respect to qubit conflict and crosstalk constraints. Two colors are assigned. (d) Either color's gate set can be selected. Assuming we choose SWAP Q0, q1 and SWAP Q3, Q4, the generated circuit is provided. And there is no crosstalk with optimal depth.  }
  \label{fig:colorgeneral}
\end{figure*}

In this section, we introduce our crosstalk-aware compilation framework. We first define the  input and output of our compiler framework, then demonstrate the major components in our framework. 

\subsection{ Input and Output}
 Our compiler framework CQC either takes a logical circuit (where there is only one way to synthesize the circuit from IR) or an IR as input. It also takes the hardware coupling graph as well as the crosstalk information as input. The crosstalk parameters can be profiled or provided by the vendor. Since we do not focus on the profiling, we adopt previous approach for profiling if necessary \cite{murali+:asplos19}.  With the consideration of physical connectivity constraint, crosstalk, and qubit decoherence effect, our framework aims to produce a compiled circuit with maximal fidelity.

\subsection{Overall Design}

 Our ultimate goal is to compile the given circuit with maximal fidelity. Ideally, the compiled circuit has minimal circuit depth and zero crosstalk. However, it may not be possible to have both at the same time.  Increasing the allowance of crosstalk, depth of compiled circuit decreases and the decoherence error decreases. Zero crosstalk allowance would make the circuit depth large which would impair the fidelity. With a reasonable tolerance of crosstalk, we may find the balance between circuit depth and crosstalk and achieve the maximal fidelity. 

 We first determine the range of crosstalk allowance. Next we choose the circuit version with the maximum fidelity in this range.

\paragraph{Maximum crosstalk allowance $X\_Max$.} We need to set the maximum crosstalk allowance. With all crosstalk allowed, we compile the input circuit to achieve maximal parallelism and minimal circuit depth. Then we calculate the amount of crosstalk in this compiled circuit and set this as our maximum crosstalk allowance $X\_Max$. The crosstalk could not be worse than that  for this version.

\paragraph{Minimum crosstalk allowance $X\_Min$.}
We set $X\_Min$ = 0. 
 
\paragraph{How to determine the best allowance X. } We use binary search to find the best crosstalk allowance in the range ($X\_Min$, $X\_Max$). The criteria is the fidelity of generated circuit.  

Given the crosstalk allowance range $X \in [X\_Min, X\_Max]$, we set the initial allowance equal to $X = (X\_Min + X\_Max)/2$. Then we compile circuit with crosstalk allowance X and X + 1 (or X and X-1).  We obtain the fidelity of two compiled circuits by calculating the  estimated Success
Probability (ESP) \cite{nishio+:JETC2020} considering both decoherence error and crosstalk errors. We can also use the fidelity measured as TVD by running the compiling circuit on real machines, if such overhead is allowed.

Then we reset $X\_Max$ = X if X+1's corresponding circuit has worse fidelity, else $X\_Min$ = X. Repeating this process until we find the best crosstalk allowance. Once we find the best crosstalk allowance, we also found the circuit with the maximum fidelity. The whole process is shown in Fig. \ref{fig:frame} (left side).

\subsection{Compile with Respect to A Given Crosstalk Allowance }

Now, Let's discuss how to compile a given circuit with fixed crosstalk allowance and crosstalk parameter. The whole process is shown in right component in Fig. \ref{fig:frame} (right side). To compile the input circuit, we go  through the circuit and save all un-executed gates in a list $Unsched\_list$ (in Algorithm \ref{alg:hs}. 

We also construct a graph called candidate set graph (CSG) where each node is a gate that can be scheduled at the current time step, where CSG = $\{V, E\}$.

$V$ consists of two types of gates. One type is the original gates in the circuit that are executable, satisfying both dependence constraints and hardware coupling constraints. The other type is SWAP gates. They consist of the SWAP gates that can help non-executable (dependence resolved) gates over overcome the physical constraint. We only focus on the SWAPs that can help reduce the distance by 1 for any pair of qubits in an un-executable CX. 

 If two vertices $v_i$ and $v_j$ share a qubit or there is a crosstalk (that is not in crosstalk allowance) between them, we add an edge ($v_i, v_j$) into set E.  
 
 \paragraph{The Chromatic Method}
 We apply a chromatic method for determining which gates to run and which SWAPs to insert.  The idea is that if two gates share a qubit (they conflict), they cannot run simultaneously; If two gates have crosstalk and the crosstalk is not in allowance, they cannot run simultaneously. The coloring method ensures that any two vertices that are adjacent cannot be given the same color, corresponding to the aforementioned constraints.  An example on how our compiler automatically found the solution in the motivating example (Fig. \ref{fig:runthroughexample}) is shown in Fig. \ref{fig:colorgeneral}.

 After coloring, all gates corresponding to a specific color can run at the same time. We pick a gate set that correspond to a color with certain criteria such as Cgate count, criticality of selected gates, the usage of crosstalk allowance, and etc  (details discussed in Section \ref{sec:techgeneric} and Section \ref{sec:techvqa}).

We run the selected gate set, update $Unsched\_gate$ and mapping. This process will be repeated until there is no gate in unscheduled circuit. The algorithm is described in Alg. \ref{alg:hs}.
 
 \paragraph{Application-specific Optimization}
 
 The overall framework in Fig. \ref{fig:frame} and Algorithm \ref{alg:hs} work for both general applications and VQA. The differences, however, are in the implementation of  the component for generating a compiled circuit given a crosstalk allowance (named \emph{Compile\_circuit(C, arch-X, xa)} component). In particular, it is in the ``Candidate gate graph construction (CSG)" and the ``Getting a gate set" components. We describe them for general applications and VQE in details in the next two sections.

{\section{Compiling for Generic Applications}\label{sec:techgeneric}}
For generic applications, there is typically one way to generate a logical circuit from the IR. Hence the order and the gates are pre-determined before compilation.

\begin{figure}[htp]
  \centering
  \includegraphics[width=0.6\linewidth]{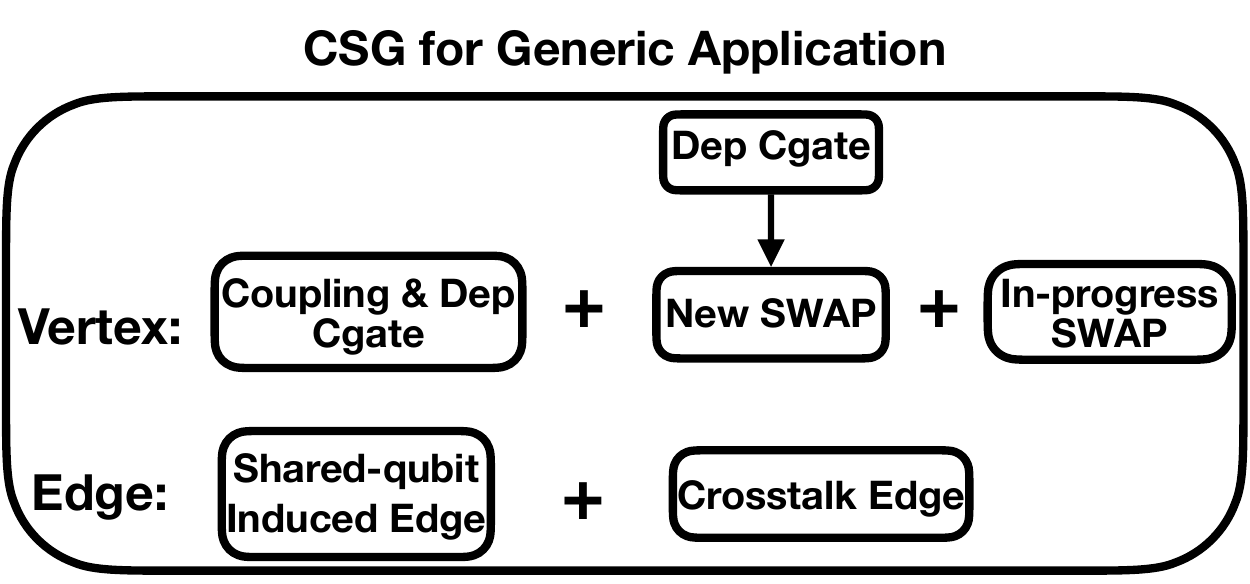}
  \caption{CSG construction for generic applications. Coupling \& Dep Cgate is coupling satisfied and dependency resolved. New SWAP gates are induced from coupling unsatisfied but dependency resolved Cgates }
  \label{fig:csg_generic_graph}
\end{figure} 

\subsection{Construct CSG For Generic Application}
\subsubsection{Determining the Vertices of CSG}
We describe how to construct candidate set graph (CSG) for generic applications. The outline is in Fig. \ref{fig:csg_generic_graph}.

A vertex in CSG corresponds to a gate. There are two types of vertices in CSG, the original gates in the circuit, which we name as \emph{circuit gate (Cgate)}, and SWAP gates that are to be inserted to the circuit.

\paragraph{Circuit gate vertex } Generic circuit imposes dependences between gates. The circuit gates that have their dependence resolved, i.e., their predecessors have been executed, and satisfied the coupling constraints, i.e., the two qubits that participate in the gate are placed in adjacent physical qubits, are considered as  gate vertices in CSG.

\paragraph{SWAP gate vertex} We choose the SWAP gates that can help the circuit gates which have their dependency but not coupling constraints resolved. We choose the SWAP gates that can reduce the distance by one for any pair of such qubits.

See an example of Cgate and SWAP vertices in Fig. \ref{fig:colorgeneral} (b). 

\paragraph{In-progress SWAP} It is note worthy that the execution time of SWAP gate may be decomposed into three CX gates (in some architectures). Hence its latency is different from the CX gate or  other two-qubit gates. 

It is possible that as the hardware mapper process the next set of gates in the frontier (dependence-resolved gates) in the iterative process (in Alg. \ref{alg:hs}), the SWAP gates from the last iteration(s) are still in progress. Those SWAP cannot be interrupted and should not let their qubits be used by any other new SWAPs. So those in-progress SWAP gates would be added as SWAP gate vertices to CSG as well.

\subsubsection{Determining the Edges of CSG}
There are two types of edges in the CSG. One type is induced from the gates that share qubits, for instance, different candidate SWAPs. The other type is induced from the crosstalk constraint.

\paragraph{Shared-qubit induced edges} If two gates(vertices) share a qubit, we add an edge connecting these two vertices. 

\paragraph{Crosstalk edge} If there exist a crosstalk between two gates, we add an edge connecting these two gates(vertices) in the CSG. 
In the case of crosstalk allowance, not all crosstalk edges will be included in the final CSG, depending on how much crosstalk is allowed at the current iteration{.  We first sort all these crosstalk edges according to their crosstalk errors in the ascending order. Then we remove a corresponding number of crosstalk edges that have least crosstalk errors. The number of edges we remove is based on how much crosstalk allowance is left at this current iteration. The crosstalk allowance is updated iteration by iteration.  }

See an example in Fig. \ref{fig:colorgeneral} (c) for edges  in CSG.

\subsection{Gate set generation by graph coloring}
 
To perform graph coloring, we use the \emph{Welsh Powell} algorithm \cite{welsh+:1967}. It sorts all the vertices by degree and color the highest degree first. It is an efficient greedy algorithm that work for large number of vertices. We slightly modified it to accommodate the case of in-progress SWAPs. If a SWAP gate is decomposed into  three CX gates, we do not want to interrupt these in-progress swaps. We first give these in-progress SWAP gates the same color, then we follow the same steps as that in Welsh Powell.

\subsection{Gate Set Ranking}
After coloring, we must select a proper color which corresponds to a particular gate set to be executed. We can select the color that corresponds to the largest number of gates. However, to maximize fidelity and minimize depth, we must consider the following three factors:

\begin{itemize}
    \item Circuit gate count.
    \item Gate criticality.
    \item Consecutive SWAPs making progress.
    \item Crosstalk allowance.
\end{itemize}

For circuit gates count, a color might correspond to a large gate set. But some of them might be just SWAP gates. We need to have a good balance between the number of SWAP gates and circuit gates to make sure the circuit makes progress. 

For gate criticality, in each color set, some gates are in the critical path of the circuit while some others are not. Delaying the execution of those critical gates would increase the circuit depth and result in higher decoherence rate. So the more gates in a color set involving the critical path gate, the higher ranking the color has.

For the SWAP gates, since we only determine the concurrent SWAP at one layer at one time. If not properly designed, a SWAP gate that is inserted in the last iteration could be counteracted by another SWAP in the current iteration. This is harmful and prevents the circuit from making progress.

With these consideration, our color set ranking is designed as follows. First, we pick top-k colors with largest gate count. Then we rank these k sets with respect to the number of Cgates. Next we pick the one containing more SWAP gate(s) which help a CX gate already helped in the last iteration, but has not been executed due to coupling constraint. By doing this, the circuit can continue to make progress. { We also use crosstalk allowance as a tie breaker at this stage. If both gate sets are ranked the same, we choose the one that uses less crosstalk allowance. }

{\section{Compiling for VQA Applications} \label{sec:techvqa}}

\subsection{Two Categories of VQA}

We categorize VQA into two types. One type encodes 2-local qubit interaction in each Pauli-string. Different Pauli-strings can commute. Each Pauli-string corresponds to one two-qubit gate. Hence the two-qubit gates can \texttt{commute}. This is a fairly common type of VQA application. Examples are quantum approximate optimization algorithm (QAOA), Ising model, and Heisenburg model. \cite{lao+:arxiv21twoqan}. 

The second type is N-local qubit interaction (N>2), where in each Pauli-string block, more than 2 qubits interact through a tree-like structure. In this case, the flexibility in circuit (tree) synthesis can be exploited for crosstalk mitigation.

\begin{figure}[htp]
  \centering
  \includegraphics[width=0.6\linewidth]{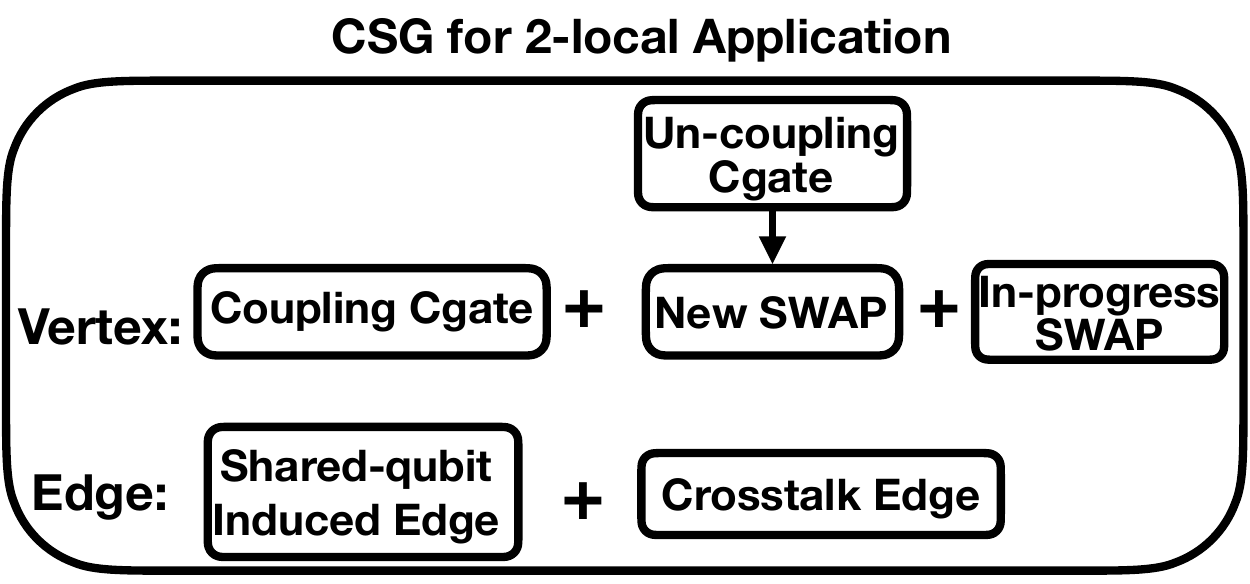}
  \caption{CSG construction for 2-local VQA. Coupling Cgate is coupling satisfied. New SWAP gate is induced from coupling unsatisfied Cgate. }
  \label{fig:qaoa_CSG}
\end{figure} 

\subsection{2-local Case}

We still use the overall framework described in Section \ref{sec:techoverview}. We in particular discuss the candidate set graph (CSG) construction (in Fig. \ref{fig:qaoa_CSG}). There is no pre-determined partial order between gates. As the two-qubit gates all commute, the vertex sets need to be constructed in a different way compared with the generic applications.

\paragraph{Circuit gate vertex:} Unlike the generic circuit, in the 2-local VQA case, the circuit gate vertex set of CSG consists of coupling-compliant and dependency-resolved gates. We look at the whole circuit and add all the coupling-compliant gates into the vertex set of CSG.

\paragraph{SWAP gate vertex:} The vertex set of CSG also contains SWAP gates that will help the coupling in-compliant gate. Similarly those SWAP gates reduce the distance between two qubits corresponding to the coupling in-compliant gates by one. In-progress SWAP gates are also included in the vertex set of CSG.

\paragraph{The edge set} The edge set is determined in  the same way as that for the generic application.

\paragraph{2-local case compilation}
Once the coloring graph constructed, we apply the graph coloring algorithm to schedule gates at the current layer as discusses in section . This process is repeated until all gates are scheduled.

\begin{figure}[htp]
  \centering
  \includegraphics[width=0.5\linewidth]{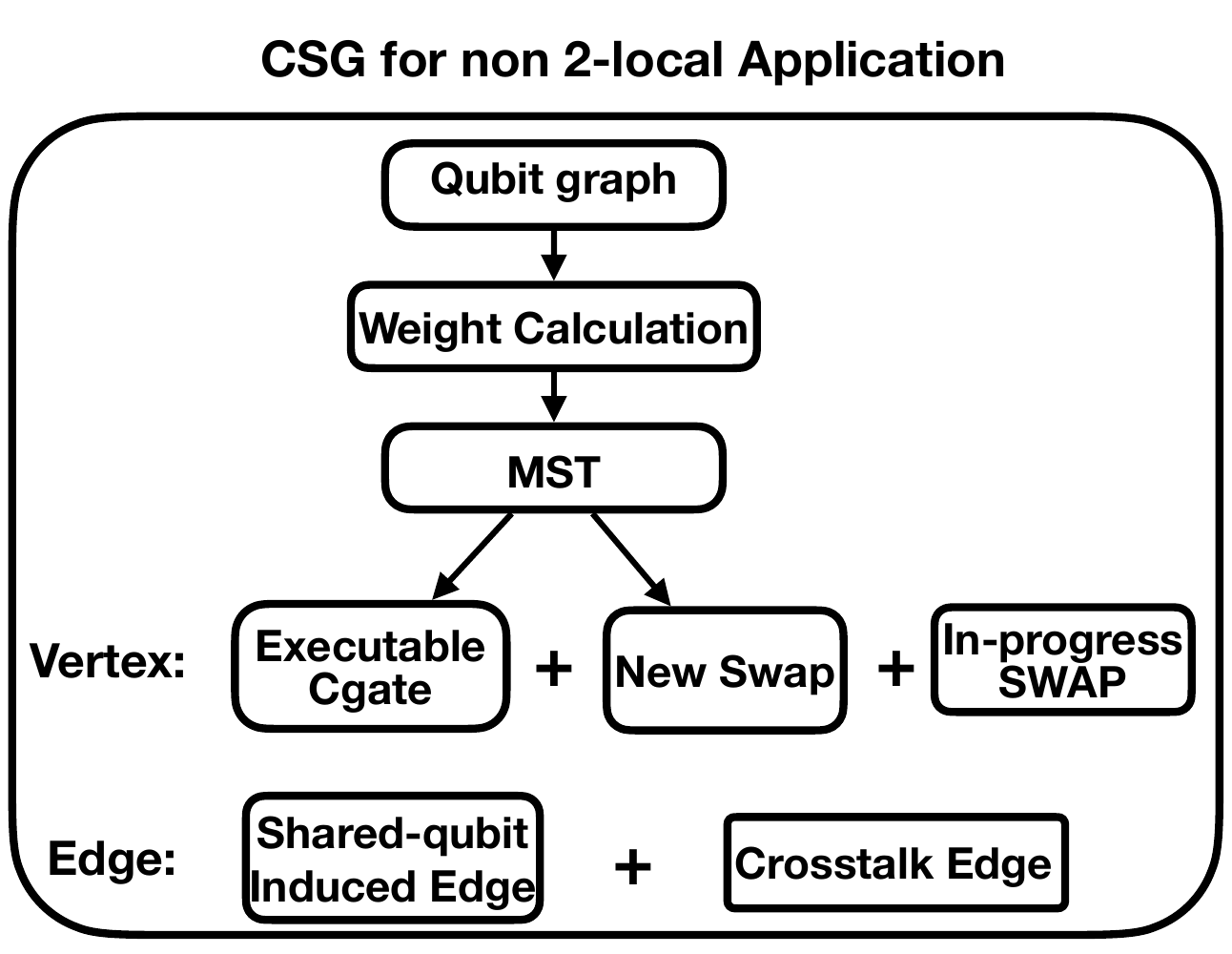}
  \caption{CSG construction for N-local VQA}
  \label{fig:vqe_frame}
\end{figure}

\subsection{N-local Cases (N$>$2)}
\subsubsection{Overview}
N-local VQA provides even more flexibility. It allows flexibly synthesis of a circuit.  Instead of decoupling the logical circuit synthesis step from the other components of crosstalk mitigation, we determine the synthesized circuit on the fly together with hardware mapping.  
The guidance for synthesizing a circuit with respect to a Pauli-string is that two-qubit gates must be applied to the qubits in the non-identity operator in Pauli-string representation. For instance, for the Pauli-string ``ZZIZZII", qubits 0, 1, 3, and 4 must be entangled through a network of CNOT gates \cite{li+:asplos22}. Note that single gates must also be inserted, but it does not affect our compilation as it is before or after the circuit block that consists of only two-qubit gates. Examples of valid synthesized circuit are shown in Fig. \ref{fig:vqe_example}. 

Now, given a Pauli-string, we must determine which two-qubit gates to execute. The rule is that there must be a root node among these qubits, and that every qubit must find a path from itself back to the root through directed edges, implemented as CX gates. The gates must be synthesized in a bottom-up manner, from leaves to the root. 

Next we describe how we synthesize the circuit (Cgates) on the fly with respect to the our candidate set graph (CSG).  

\subsubsection{The Vertex Set of CSG}
\paragraph{Qubit graph} In order to decide the vertex set of CSG for N-local VQE, we use  the minimal spanning tree (MST) method. We apply the MST method to a different graph -- the qubit graph (QG). A qubit graph is a clique graph with each node standing for a non-identity Pauli-operator qubit. However, the weight of the edges in the clique are different. 

As the example shows in Fig. \ref{fig:vqe_mst_color}(c). The input is a Pauli-string "ZZIZZII" in Fig. \ref{fig:vqe_mst_color}(a). There are four qubits that correspond to non-identify Pauli-operators. The mapping for logical qubits are shown in \ref{fig:vqe_mst_color}(b), highlighted.

\paragraph{The Weight of Edges in QG}
The weight of each edge in QG is the shortest distance between the two physical qubits where the logical qubits are mapped to. Certain edges in the qubit graph has distance of 1 while others may have larger distance. 
\paragraph{Finding MST in QG}
Based on the qubit graph, we find the minimum spanning tree (MST) from it. A minimal spanning tree is a subset of edges that connect all vertices together without any cycle, and in the meantime, have smallest total weight. For instance, in Fig. \ref{fig:vqe_mst_color} (c), the minimal spanning tree is q0-q1-q3 with total weight 4. MST tends to select the nodes that are already connected since their weight is 1, the lowest.  It is as if we give these edges of weight 1 higher priority during circuit synthesis. Since they are already executable, if we choose these edges as gates in the synthesized circuit, they do not need any SWAP gates.

\paragraph{Finding the vertex set of CSG}
From the result of MST, we can determine the gates (vertices) in CSG.  We extract three types of gate from the MST. The MST indicates three set of gates, candidate Cgates, non-executable gates, and candidate SWAP gates. 
\begin{itemize}
    \item \textbf{The Cgates set} contains CX gates corresponding to edges of weight 1 in the MST.
    \item \textbf{The non-executable gate set} contains the edges of weight greater than 1 in the MST, and the in the meantime one of the vertices of such edge must have degree 1 in the MST.  
    \item \textbf{Candidate SWAP gates set} contains SWAP gates that reduce the shortest distance between every two qubits for each gate in the non-executable gates set.
\end{itemize}

A subtlety here is that for the non-executable gate set, it is not the entire set of edges in the MST with weight > 1. Because we must follow the order of executing gates (as if edges in a tree) from a leave to the root, the edges in the middle of  the MST (edges with degree of both nodes $>=2$) are unlikely to be executed first anyway. Hence, we only choose the edges in MST which connects one end node (degree of 1).  

An example of these three sets are shown in Fig. \ref{fig:vqe_mst_color} (d). All gates in the Cgates set,  and the SWAP set are added to the vertex set of CSG. In addition, those in-progress SWAP gates are included in CSG like in generic applications.

\subsubsection{The edge set of CSG}

The edges of CSG for N-local VQE are determined the same as that for the generic application.  Fig.\ref{fig:vqe_mst_color}(e) shows the CSG graph. 

\subsubsection{Workflow Modification for the N-local Case}

The handling of the N-local case is a bit different since we need to determine which gates to synthesize at every step, and we have to ensure that we eventually synthesize a tree and that tree-induced circuit is efficient. Hence, we make a slight modification to the workflow as follows. 

 After the CSG constructed, we apply graph coloring approach in the same way as generic application compilation.

\begin{figure*}[htp]
  \centering
  \includegraphics[width=1.0\linewidth]{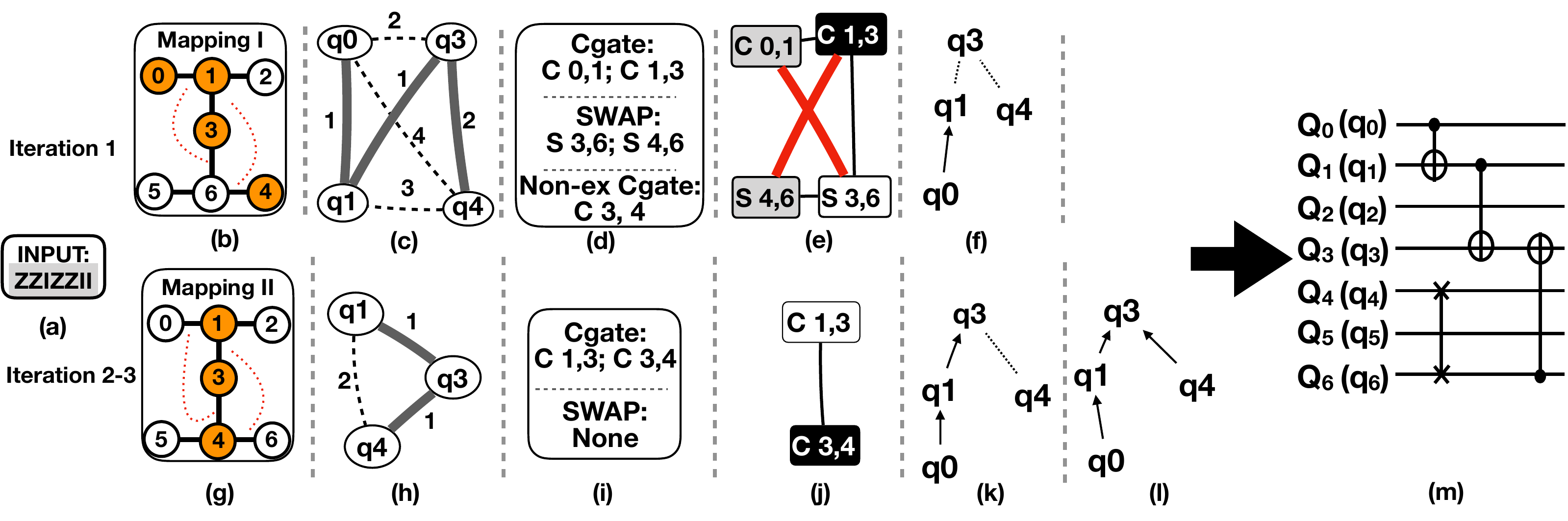}
  \caption{ N-local VQE application example. (a) Input pauli string; (b) Initial mapping. Mapping for pauli-operator qubits are highlighted in yellow (c) Qubit Graph and MST; (d)Induced Condidate Cgate set, Candidate SWAP set, and Non-executable Cgate set from (c); (e) Colored graph corresponds to (d); (f) Tree with Cgate(q0, q1) scheduled; (g) Mapping in the second iteration; (h) Qubit graph and SMT for remaining pauli string; (i) Induced Condidate Cgate set and Candidate SWAP set from (h); (j) Colored graph corresponds to (i); (k) Tree with Cgate(q1, q3) scheduled; (l) The final tree; (m) The final scheduled circuit without crosstalk. }
  \label{fig:vqe_mst_color}
\end{figure*}

An example for the colored graph is shown in Fig. \ref{fig:vqe_mst_color}(e) where the gray vertex set is larger than other colored vertex set. Therefore we choose to schedule Cgate(0, 1) and SWAP(4, 6). Then we update the corresponding mapping  since a SWAP gate is applied, as shown in Fig. \ref{fig:vqe_mst_color}(g).  

Now we describe the changes to the workflow here. 

\paragraph{Gate execution. } Despite the fact that we have determined which Cgate to synthesize given the MST, we haven't determined the the control and target qubits in each selected Cgate.  It is important, since the control-target relationship determines the direction of edges in the formed tree. Either direction is okay, as long as eventually we form a tree for a node to find a path all the way to the root. But it might add SWAP cost, if the direction is set up in a un-desired way.

 In order to reduce such SWAP cost, we find the graph center of MST. We let the qubit closer to the center be the target qubit and the other qubit be the control qubit. For instance, in our example Fig. \ref{fig:vqe_mst_color} (f), we let Cgate(q0, q1) take q0 as control qubit and q1 as target qubit.

\paragraph{Qubit Deleting.} After the direction is determined, we remove the control qubit from the consideration for a future iteration since every time we remove a leaf node and there is no other qubit that points to it. Then we handle the rest of the qubits.  For example, in the second iteration of Fig. \ref{fig:vqe_mst_color}(h), the qubit graph does not contain qubit q0 anymore.  

Therefore in the workflow, we need to perform control-target direction determination and qubit deletion after a gate set is selected. It is worth mentioning that at each iteration of the major workflow component \texttt{Compile\_circuit} in Fig. \ref{fig:frame}, we repeat the same process of qubit graph construction, MST search, and so on. The MST helps us determine a part of the gates that can be executed at one iteration, but it does not mean the tree determined by MST is the final tree. Our example happens to show that MST gives the final synthesized circuit tree, but it is not necessarily the case. The qubit mapping changes over time, and the best circuit tree may change from iteration to iteration too. That's why we perform circuit synthesis on the fly.

Our example shows a complete process for solving the motivation example circuit in Fig. \ref{fig:vqe_example}  in Fig.\ref{fig:vqe_mst_color}. The last few steps (h) to (l) show how the rest of the circuit is handled. 

\section{Evaluation}
\label{sec:eval}
In this section, we evaluate CQC by comparing it with  the state-of-the-art. First we introduce our experiment setup including the backend, metric, and baselines. Then we show the fidelity and performance (depth) of CQC compared with prior approaches. We perform experiments on both real machine and simulator. For any benchmark with less than 20 qubits, we run it on real machine. Beyond 20 qubits, we run it using simulation. Both generic applications and VQA applications are included for evaluation. 

\subsection{Experiment setup}
\paragraph{{Backend}} All real-machine experiments are performed  on the machine, IBMQ 27-Qubit IBM-Geneva.

 For the experiments of large benchmarks with more than 20 qubits, we use the IBM Washington simulator.  IBM Washington is the largest near-term 127-qubit architecture. The classical backend machine for simulation has Intel Xeon CPU E5-1607 with  4 cores at 3 GHz.

 Since IBM Qiskit simulator does not automatically take crosstalk error into account, we modified it to reflect the crosstalk errors between different links of qubits.    Although the IBM Washington simulator can support up to 127 qubits, due to the classical backend machine's memory and speed limit, we only simulate the benchmarks for qubit number up to 30 qubits and gate number up to 50000.

\paragraph{{Metric}} We use two metric: total variant distance (TVD) \cite{Nielsen+:2002book, knill+:quant-ph1995} and depth. TVD measures the fidelity of the generated circuit. The minimum distance TVD of 0 is achieved when the compiled circuit runs with zero error. The lower the TVD is, the more promising the compiled circuit is.  Depth measures the performance of the compiled circuit, that is, how long it takes to run.  For depth, we assume each gate takes a unit of time and a SWAP gate consists of  three CX gates. Note that the depth not only contributes to the performance, but also the decoherence error. A smaller depth circuit tends to have smaller decoherence errors. Therefore, if other error rates are the same, the smaller the depth is, the better.  

\paragraph{{Benchmarks}} 
The benchmarks used for evaluation are selected from RevLib~\cite{Wille+:ISMVL08}, IBM Qiskit~\cite{Qiskit}, QASMBench~\cite{li+:2021qasm},  and ScaffCC~\cite{JavadiAbhari+:CCF14}. We show the qubit and gate number in Table.  \ref{tab:result}. 

There are three types of benchmarks. The first type is the generic benchmark. The second type is for 2-local VQA application. We use QAOA \cite{farhi+2014quantum} for the 2-local VQA benchmark. For QAOA we choose different input-problem graphs ranging from 5 vertices to 30 vertices. The gate number is the same as the edge number in the prolem graph. We show the gate number in Table \ref{tab:result} as well. 

The third type is N-local VQA, for which we choose Hamiltonian simulation. These are variational quantum eigensolvers (VQE) generated using Pyscf~\cite{sun+:2018cms}. We choose four different
molecules (H2, Li-H, Li-Li, H4) with different basis (sto3g, mingao) to generate Hamiltonian functions. There are six benchmarks in this category. 

\paragraph{{Baseline}}
For generic applications and QAOA, we use the best-known approach called Xtalk \cite{murali+:asplos19} as our baseline. Xtalk optimizes crosstalk at scheduling level. It uses SMT solver to determine where and if barriers should be added to delay the execution. For hardware mapping, we use Sabre \cite{li+:asplos19} which is an efficient mapper integrated in Qiskit. We name this baseline as ``Xtalk".

For VQE applications, we use Paulihedral \cite{li+:asplos22} as our baseline, which is the state-of-art approach to synthesize circuits from Pauli-string specifications.

%And to suit larger benchmarks, We use IBM's 5-qubit Santiago and Bogota 14-qubit Melbourne, 16-qubit Guadalupe, 20-qubit Q20 Tokyo architecture \cite{li+:asplos19} and 27-qubit Montreal , which has different connectivity and our method can scale for different kind of architecture which are shown in Figure~\ref{fig:IBM-machine}. Some machines like Tokyo and Melbourne represent different topology, although they are retired, they are still researcher target by existing works\cite{li2022paulihedral}\cite{murali+:asplos19}\cite{ostaszewski2021structure}

\subsection{Experiment result}
 
We show the overall experiment results in Table \ref{tab:result}. We have  improvements in fidelity compared with prior work, ranging from 5\% to 23\%, and depth improvement from 7\% to 39\%. 

% \begin{figure}
% \centering
% \begin{subfigure}{.5\textwidth}
%   \centering
%   \includegraphics[width=.4\linewidth]{fig/generic_circuit.pdf}
%   \caption{Generic application results. X-axis shows benchmark names. Y-axis shows TVD.}
%   \label{fig:generic result}
% \end{subfigure}%
% \begin{subfigure}{.5\textwidth}
%   \centering
%   \includegraphics[width=.4\linewidth]{fig/qaoa.pdf}
%   \caption{QAOA TVD results. X-axis shows QAOA benchmark names. Y-axis shows TVD.}
%   \label{fig:QAOAresult}
% \end{subfigure}

% \end{figure}

\begin{figure}[!t]
  \centering
  \includegraphics[width=0.4\linewidth]{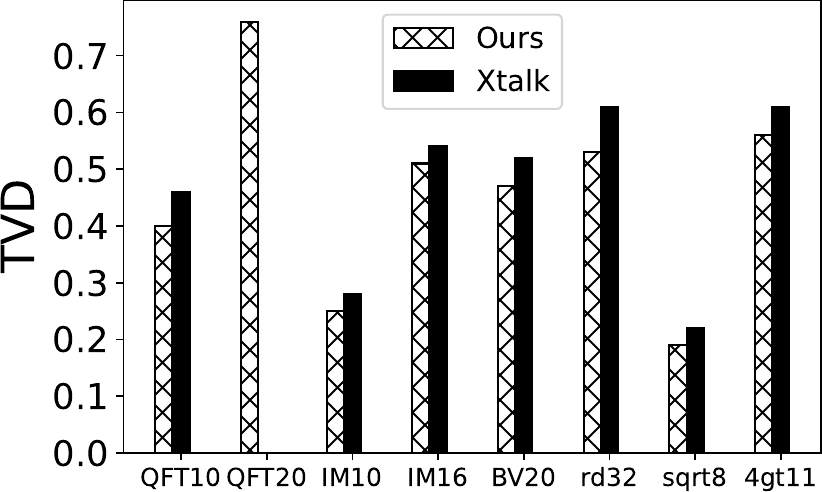}
  \caption{Generic application results. X-axis shows benchmark names. Y-axis shows TVD.}
  \label{fig:generic result}
\end{figure} 

\begin{figure}[!t]
  \centering
  \includegraphics[width=0.4\linewidth]{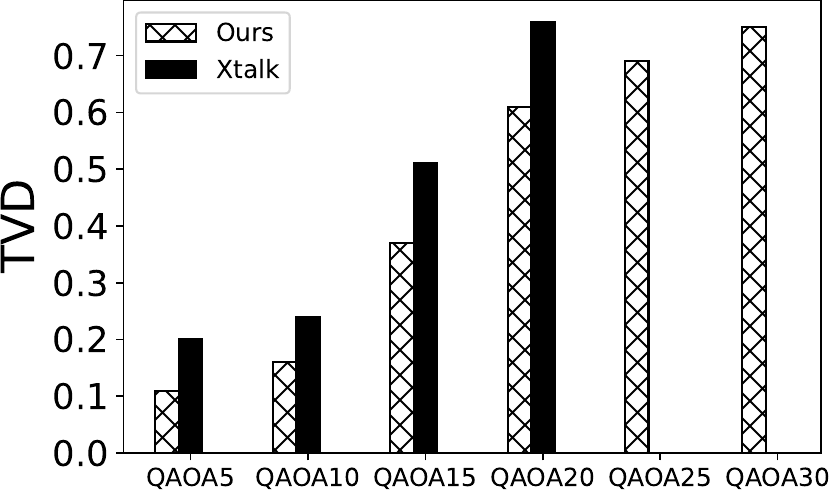}
  \caption{QAOA TVD results. X-axis shows QAOA benchmark names. Y-axis shows TVD.}
  \label{fig:QAOAresult}
\end{figure} 

\begin{figure}[!t]
  \centering
  \includegraphics[width=0.4\linewidth]{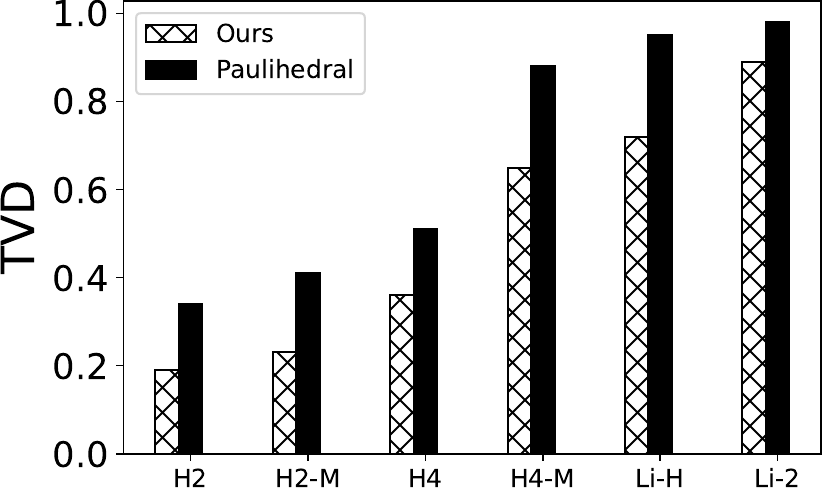}
  \caption{VQE TVD results. The X-axis shows different molecular under different basis.}
  \label{fig:vqeresults}
\end{figure}

\paragraph{Generic applications} We first show the TVD results for generic applications in Fig. \ref{fig:generic result}. We have consistently better TVD compared with that generated by Xtalk. On average we improved TVD by 5\% and at maximum by 15\% for benchmarks like {QFT} where there are more gates. 

 From depths in Table \ref{tab:result}, we can see further improvement, where  we reduce the depth by up to 39\% for the 4GT11 benchmark. It is mainly because instead of only adding a barrier for crosstalk mitigation, we exploit the hardware mapping opportunities. 

Moreover, it is worth mentioning that the SMT solver may not be able to run  large-scale circuit. For example, for the QFT20 benchmark, it can not return a result within a reasonable amount of time, for instance, 1 hour, while our method can compile the QFT20 circuit and generate a reasonable compiled circuit. 

\begin{figure}[!ht]
  \centering
  \includegraphics[width=0.5\linewidth]{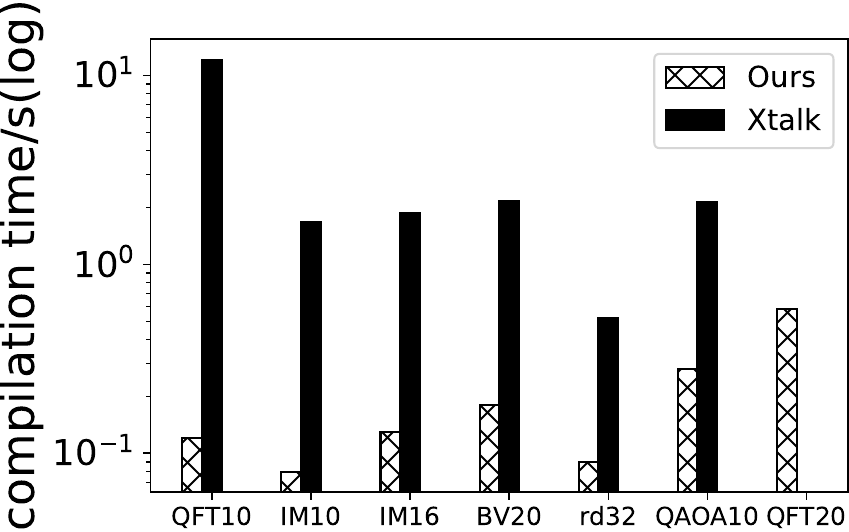}
  \caption{Compilation time results : We compare our compilation time with Xtalk, the X-axis lists different benchmark we use, we show the time (seconds) in log scale  on the Y-axis.}
  \label{fig:compilation_result}
\end{figure}

We show a comparison of compilation time in Fig. \ref{fig:compilation_result}. Our compiler takes 1\% of the prior work's compilation time in most cases. The scalability of the compiler is also important, especially current quantum machines are expected to scale beyond 1,000 qubits in a couple of years.

\begin{table}[]
\resizebox{.6\columnwidth }{!}{
\begin{tabular}{|ccc|cc|cc|}
  
\hline
\multicolumn{3}{|c|}{\textbf{Benchmark}}                                                                               & \multicolumn{2}{c|}{CQC}          & \multicolumn{2}{c|}{Xtalk/Pauli-hedral} \\ \hline
\multicolumn{1}{|c|}{Generic}  & \multicolumn{1}{c|}{Qubits} & Gates                                                   & \multicolumn{1}{c|}{Depth} & TVD  & \multicolumn{1}{c|}{Depth}    & TVD     \\ \hline
\multicolumn{1}{|c|}{QFT10}    & \multicolumn{1}{c|}{10}     & 90                                                      & \multicolumn{1}{c|}{76}    & 0.40 & \multicolumn{1}{c|}{81}       & 0.46    \\ \hline
\multicolumn{1}{|c|}{QFT20}    & \multicolumn{1}{c|}{20}     & 380                                                     & \multicolumn{1}{c|}{181}   & 0.76 & \multicolumn{1}{c|}{-}       & -      \\ \hline
\multicolumn{1}{|c|}{IM10}     & \multicolumn{1}{c|}{10}     & 80                                                      & \multicolumn{1}{c|}{72}    & 0.25 & \multicolumn{1}{c|}{80}       & 0.28    \\ \hline
\multicolumn{1}{|c|}{IM16}     & \multicolumn{1}{c|}{16}     & 150                                                     & \multicolumn{1}{c|}{128}   & 0.51 & \multicolumn{1}{c|}{133}      & 0.54    \\ \hline
\multicolumn{1}{|c|}{BV10}     & \multicolumn{1}{c|}{10}     & 9                                                       & \multicolumn{1}{c|}{14}    & 0.22 & \multicolumn{1}{c|}{19}       & 0.25    \\ \hline
\multicolumn{1}{|c|}{BV20}     & \multicolumn{1}{c|}{20}     & 19                                                      & \multicolumn{1}{c|}{30}    & 0.47 & \multicolumn{1}{c|}{32}       & 0.52    \\ \hline
\multicolumn{1}{|c|}{RD32}     & \multicolumn{1}{c|}{8}      & 51                                                      & \multicolumn{1}{c|}{51}    & 0.53 & \multicolumn{1}{c|}{64}       & 0.61    \\ \hline
\multicolumn{1}{|c|}{4GT11}    & \multicolumn{1}{c|}{5}      & 18                                                      & \multicolumn{1}{c|}{23}    & 0.19 & \multicolumn{1}{c|}{32}       & 0.22    \\ \hline
\multicolumn{1}{|c|}{SQRT}     & \multicolumn{1}{c|}{12}     & 1120                                                    & \multicolumn{1}{c|}{791}   & 0.56 & \multicolumn{1}{c|}{820}      & 0.61    \\ \hline
\multicolumn{1}{|c|}{QAOA}     & \multicolumn{1}{c|}{Qubits} & edges                                                   & \multicolumn{1}{c|}{Depth} & TVD  & \multicolumn{1}{c|}{Depth}    & TVD     \\ \hline
\multicolumn{1}{|c|}{QAOA5}    & \multicolumn{1}{c|}{5}      & 10                                                      & \multicolumn{1}{c|}{12}    & 0.11 & \multicolumn{1}{c|}{17}       & 0.20    \\ \hline
\multicolumn{1}{|c|}{QAOA10}   & \multicolumn{1}{c|}{10}     & 38                                                      & \multicolumn{1}{c|}{107}   & 0.16 & \multicolumn{1}{c|}{129}      & 0.24    \\ \hline
\multicolumn{1}{|c|}{QAOA15}   & \multicolumn{1}{c|}{15}     & 80                                                      & \multicolumn{1}{c|}{216}   & 0.37 & \multicolumn{1}{c|}{259}      & 0.51    \\ \hline
\multicolumn{1}{|c|}{QAOA20}   & \multicolumn{1}{c|}{20}     & 110                                                     & \multicolumn{1}{c|}{286}   & 0.61 & \multicolumn{1}{c|}{410}      & 0.76    \\ \hline
\multicolumn{1}{|c|}{QAOA25}   & \multicolumn{1}{c|}{25}     & 130                                                     & \multicolumn{1}{c|}{316}   & 0.69 & \multicolumn{1}{c|}{-}       & -      \\ \hline
\multicolumn{1}{|c|}{QAOA30}   & \multicolumn{1}{c|}{30}     & 150                                                     & \multicolumn{1}{c|}{428}   & 0.75 & \multicolumn{1}{c|}{-}       & -     \\ \hline
\multicolumn{1}{|c|}{VQE}      & \multicolumn{1}{c|}{Qubits} & \begin{tabular}[c]{@{}c@{}}Pauli\\ strings\end{tabular} & \multicolumn{1}{c|}{Depth} & TVD  & \multicolumn{1}{c|}{Depth}    & TVD     \\ \hline
\multicolumn{1}{|c|}{H2}       & \multicolumn{1}{c|}{4}      & 15                                                      & \multicolumn{1}{c|}{40}    & 0.19 & \multicolumn{1}{c|}{60}       & 0.34    \\ \hline
\multicolumn{1}{|c|}{H2-minao} & \multicolumn{1}{c|}{4}      & 25                                                      & \multicolumn{1}{c|}{72}    & 0.23 & \multicolumn{1}{c|}{104}      & 0.41    \\ \hline
\multicolumn{1}{|c|}{H4}       & \multicolumn{1}{c|}{8}      & 316                                                     & \multicolumn{1}{c|}{390}   & 0.36 & \multicolumn{1}{c|}{691}      & 0.51    \\ \hline
\multicolumn{1}{|c|}{H4-minao}       & \multicolumn{1}{c|}{8}      & 362                                                     & \multicolumn{1}{c|}{1381}  & 0.65 & \multicolumn{1}{c|}{2293}     & 0.88    \\ \hline
\multicolumn{1}{|c|}{Li-H}     & \multicolumn{1}{c|}{12}     & 631                                                     & \multicolumn{1}{c|}{3117}  & 0.72 & \multicolumn{1}{c|}{4709}     & 0.95    \\ \hline
\multicolumn{1}{|c|}{Li2}      & \multicolumn{1}{c|}{16}     & 2980                                                    & \multicolumn{1}{c|}{19850} & 0.89 & \multicolumn{1}{c|}{25856}     & 0.98    \\ \hline

\end{tabular}
}
\caption{Overview of Experimental Results}
\label{tab:result}
\end{table}

\paragraph{Variational Quantum Algorithms }
We show TVD results for QAOA in Fig. \ref{fig:QAOAresult}, and for VQE in Fig. \ref{fig:vqeresults}.  For QAOA, we have the TVD improvement ranging from 10\% to 15\%. Since Xtalk failed to compile for the 25-qubit and 30-qubit case due to the compilation time overhead, we do not have TVD result for Xtalk in these two cases.

For VQE, we find even more TVD and depth improvement compared with the state-of-the-art. It is because for VQE, we have  more optimization opportunities at circuit synthesis stage.  

For VQE, we have TVD improvement  ranging from 15\% to 23\%. And we can also find that CQC takes  60\% depth compared with Paulihedral in some benchmarks and 69\% on average depth. 

From the results, we can see that the compiled quantum circuit's fidelity (calculated as 1-TVD) drops below 30\% when we have QAOA with 30 vertices and 150 edges, or Li2 with 26 qubit and 2980 Pauli strings. However, Paulihedral's fidelity drops to 5\%. That means our work outperforms the state-of-the art by improving fidelity by 6X.

\section{Related work}
  \label{sec:related}
%grammarly tools to check it? 

A vast body of prior work exists on physical and software have been proposed to mitigate the crosstalk problem. There are mainly two categories of studies which focus on different parts: 1) Physical level mitigation: connectivity reduction \cite{ding+:micro20} or frequency tuning \cite{ding+:micro20}; 2) The other study focus on the software level which is to delay gates \cite{murali+:asplos19} or to perform pulse-level optimization \cite{xie+:asplos22}. The first category focuses on physical level which has strict physical requirements, the methods might not suit  all the machines. For example some machine architecture have been decided and there are some extra costs when tune the frequency of the qubits \cite{ware2010architecting}. The second category is at the software level, although delaying gate seems promising to reduce the crosstalk-error \cite{murali+:asplos19}, it might increase the de-coherence error \cite{shor+:pra95} due to longer execution depths. 
  
  However, all prior work takes it for granted that existing hardware mapping methods \cite{murali2019noise,li+:asplos19}  or circuit synthesis methods are good enough without considering crosstalk mitigation.  So they are doing the crosstalk mitigation all after the mapping and inserting swaps process. However, far too little attention has been paid to take the crosstalk-error into consideration for all mapping and insert swaps method, which leaves a new work for a new crosstalk-aware mapping strategy which take crosstalk into consideration when they map the logical qubits to physical qubits and inserting swaps. Although a lot of work exists for mapping logical qubit to physical qubit and inserting swaps, like (Sabre \cite{li+:asplos19}, the one by Zulhner \etal  \cite{Zulehner+:DATE18}). These methods are considering minimizing the decoherence error by decreasing the number of inserting swaps instead of considering crosstalk. So we are first to build a new crosstalk-aware mapping framework which minimize the crosstalk error in the mapping process and circuit synthesis stage, which can have a significant advantage compared to the prior work since a good mapping process or synthesized circuit can already mitigate or completely remove  the crosstalk errors which can eventually decrease the depths and decoherence error.

To the best of our knowledge, this is the first work to build a crosstalk-aware mapping method which consider the crosstalk error and have a better performance(error rate/depths) compared to prior work. It first considers mitigating crosstalk error together with schedule gates/inserting swaps and mapping. Also our work first use  commutative to schedule gates to avoid crosstalk error without any extra cost. 

\section{Conclusion}
Crosstalk remains one of the major contributing factors for the high error rate of contemporary superconducting NISQ devices. Existing studies consider crosstalk at the scheduling stage or pulse generation stage compilation, leading to sub-optimal performance and/or fidelity. This paper presents the first crosstalk-aware quantum program compilation framework CQC that comprehensively addresses circuit synthesis, qubit routing, and crosstalk effect altogether during the compilation. Evaluations on real NISQ machines demonstrate up to 23\% error rate reduction with 60\% circuit depth compared to state-of-the-art approaches.
\label{sec:conclusion}
% no \IEEEPARstart

\section*{Acknowledgement}
This material is based upon work partially supported by the U.S. Department of Energy, Office of Science, National Quantum Information Science Research Centers, Co-design Center for Quantum Advantage (C2QA) under contract number DE-SC0012704. We would like to thank the Pacific Northwest National Laboratory operated IBM-Q Hub. The Pacific Northwest National Laboratory is operated by Battelle for the U.S. Department of Energy under Contract DE-AC05-76RL01830.
This work was supported by the "Embedding QC into Many-body Frameworks for Strongly Correlated Molecular and Materials Systems'' project, which is funded by the U.S. Department of Energy, Office of Science, Office of Basic Energy Sciences (BES), the Division of Chemical Sciences, Geosciences, and Biosciences (under award 72689).

\bibliographystyle{plain}
\bibliography{references}
\end{document}